\documentclass{article}
\usepackage{graphicx} 
\usepackage{amsmath}
\usepackage{amssymb}
\usepackage{bbm}
\usepackage{mathtools}
\usepackage{siunitx}
\usepackage{appendix}
\usepackage[most]{tcolorbox}
\usepackage[bb=boondox]{mathalfa}
\usepackage{cite}
\usepackage{enumitem}
\usepackage{xcolor} 
\usepackage{soul}
\usepackage{geometry}
\usepackage{authblk}
\usepackage{jcappub}
\hypersetup{
    colorlinks=true,       
    linkcolor=black,        
    citecolor=black,        
    urlcolor=black          
}
\usepackage{lineno}

\DeclarePairedDelimiter\bra{\langle}{\rvert}
\DeclarePairedDelimiter\ket{\lvert}{\rangle}
\DeclarePairedDelimiterX\braket[2]{\langle}{\rangle}{#1\,\delimsize\vert\,\mathopen{}#2}

\newcommand{\bs}[1]{\mathbf{#1}}
\newcommand{\al}{\alpha}

\title{Many-body Response to Spin-Independent and Spin-Dependent Dark Matter Scattering in a Superfluid Helium-3 Target
}
 \author[a]{\large QUEST-DMC Collaboration: A. Ting}
 \author[a]{N. Darvishi,}
 \author[b]{S. Autti,}
 \author[c]{L. Bloomfield,}
 \author[a]{A. Casey,}
 \author[a]{N. Eng,}
 \author[c]{P. Franchini,}
 \author[b]{R. P. Haley,}
 \author[a]{P. J. Heikkinen,}
 \author[e]{A. Jennings,}
 \author[f]{A. Kemp,}
 \author[c,a]{E. Leason,}
 \author[c]{J. March-Russell,}
 \author[c]{J. Monroe,}
 \author[b]{D. M{\"u}nstermann,}
 \author[b]{M. T. Noble,}
 \author[b]{J. R. Prance,}
 \author[a]{X. Rojas,}
 \author[b]{T. Salmon,}
 \author[a]{J. Saunders,}
 \author[d]{J. Smirnov,}
 \author[a,c]{R. Smith,}
 \author[b]{M. D. Thompson,}
 \author[b]{A. Thomson,}
 \author[b]{V. Tsepelin,}
 \author[a]{S. M. West,}
 \author[b]{L. Whitehead,}
 \author[b]{D. E. Zmeev}

 \affiliation[a]{\textit{Department of Physics, Royal Holloway University of London, Egham, Surrey, TW20 0EX, UK}}
\affiliation[b]{\textit{Department of Physics, Lancaster University, Lancaster, LA1 4YB, UK}}
\affiliation[c]{\textit{Department of Physics, University of Oxford, Keble Road, Oxford, OX1 3RH, UK}}
\affiliation[d]{\textit{Department of Mathematical Sciences, University of Liverpool, Liverpool, L69 7ZL, UK}}
\affiliation[e]{\textit{RIKEN Center for Quantum Computing, RIKEN, Wako, 351-0198, Japan}}
\affiliation[f]{\textit{UKRI STFC Rutherford Appleton Laboratory, Particle Physics Department, Harwell, Didcot OX11 0QX, UK}}

\emailAdd{adam.ting@rhul.ac.uk}\emailAdd{neda.darvishi@rhul.ac.uk}

\date{}

\abstract{\noindent The QUEST-DMC experiment employs an ultra-low-temperature superfluid $^3$He target in an effort to detect sub-GeV dark matter (DM). The low-momentum scattering of such light DM is sensitive to the many-body structure of superfluid $^3$He, making it necessary to account for the collective behaviour of the medium. We present a first-principles calculation of the DM scattering rate in this inherently quantum many-body medium for both spin-independent and spin-dependent interactions. The leading consequence of this many-body behaviour is a kinematic enhancement of the scattering rate arising from the Fermi–Dirac statistics of the ultra-low-temperature target. Accounting for this extends the projected reach to lower DM masses for both spin-independent and spin-dependent interactions under the benchmark detector assumptions considered here.
~The resulting sensitivity improvement becomes increasingly pronounced as the detector threshold drops into the sub-eV regime. It is negligible at the conventional readout threshold of approximately $31$~eV, but becomes significant at the $0.51$~eV benchmark threshold for SQUID-based readout. Specifically, at this threshold, incorporating the many-body response extends the higher-energy recoil tail and improves the projected low-mass reach by approximately $2.5\%$ relative to the single-atom approximation. This improvement becomes more pronounced at lower thresholds. Reaching a threshold of $5$~meV could further extend the sensitivity of QUEST-DMC into the realm of sub-MeV DM.
}

\begin{document}

\maketitle

\section{Introduction}

There is overwhelming evidence across a vast range of astronomical scales that the majority of the matter content of our universe is non-luminous, non-baryonic dark matter (DM). This evidence probes the gravitational effects of DM but does little to clarify its other properties. Historically, many direct- and indirect-detection searches have focused on weak-scale DM candidates with masses in the range 10–1000 GeV/c$^2$. These searches have been extensive but have yielded no confirmed detection so far, providing impetus to expand the search to a broad class of theoretically well-motivated lighter DM particles with masses below that of the proton~\cite{Hall_2010,Barnes_2020,Hochberg_2014,Zurek_2014,Pospelov_2008,Jaeckel:2012mjv, boehm2004light, boehm2004scalar, pospelov2008secluded}.

For sub-GeV DM candidates, direct detection of DM-nucleus scattering events requires experiments with low-energy thresholds capable of detecting recoil energies
below the keV scale.
Earlier work~\cite{quest2024quest,darvishi2025dark,darvishi2025attenuation,autti2023long,autti2024quest,leason2026development} introduced the Quantum Enhanced Superfluid Technologies for Dark Matter and Cosmology (QUEST-DMC) direct detection experiment. Its $^3$He target is operated at temperatures below 200 $\si{\micro \K}$ where it exists in a superfluid quantum phase. 
Compared to other noble gases, the light $^3$He nucleus is kinematically well matched to sub-GeV masses for maximising recoil energy, while additionally giving the experiment sensitivity to both spin-independent (SI) and spin-dependent (SD) interactions through its spin-$\tfrac{1}{2}$ nature. This $^3$He target therefore allows QUEST-DMC to search for a wealth of light DM models. The experiment projects operation with a sub-eV energy threshold, in principle providing it sensitivity to DM with masses below 25~MeV/c$^2$. 

Previous work~\cite{quest2024quest,darvishi2025dark,darvishi2025attenuation} considered the DM signal calculation in the regime where the individual constituent $^3$He atoms are resolved, and effects of the incoherence of the nucleus, relevant at high momentum transfer, are accounted for.
The momentum of a DM particle with virial velocities has momentum corresponding to a wavelength \mbox{$\lambda\sim\si{\nm} /\left(\tfrac{m_\chi c^2}{\si{\MeV}}\right)$}. The QUEST-DMC target has a density corresponding to a few Angstrom interatomic spacing and a nanometre-scale Fermi wavelength. The scattering of 
a DM particle with a mass of a few MeV/c$^2$ can therefore resolve the many-body structure of the superfluid $^3$He target medium. 

Excitation of collective modes from DM scattering events has been extensively studied in solid state systems and the bosonic superfluid Helium-4 ($^4$He)~\cite{trickle2020multi,  kahn2022searches, gori2025spin, schutz2016detectability,hochberg2022new, hochberg2023directional, derenzo2017direct}. In this work we apply and extend established methods to describe excitation of fermionic quasiparticles due to DM scattering in superfluid $^3$He. In Section~\ref{section:ManybodyScattering} we outline the calculation from first principles of the event rate for DM scattering off a many-body target, for both benchmark SI and SD DM interactions. This coherent response captures the many-body physics relevant to the low-momentum scattering of lighter DM. 
In Section~\ref{section:ScatteringInSF3He} we present a construction of the scattering rate in the QUEST-DMC superfluid $^3$He target specifically. DM with masses greater than around $20$~MeV/c$^2$ can produce recoil energies above an eV, for which the signal for an incoherent target is effectively reproduced. We hence find the DM signal produced from this method both validates previous work by confirming current sensitivity projections of the experiment, and additionally provides the means to make the considerations of the target medium's quantum nature necessary for near-future runs. 

At sub-eV recoil energies, quantum effects become clearly apparent. This energy regime is relevant to DM masses around 10~MeV/c$^2$ and lower. With the projected benchmark energy threshold of $0.51$ eV for SQUID-based readout in QUEST-DMC, these effects become relevant to future searches for light DM. One such effect is that the quantum degeneracy pressure within the target requires finite momentum atoms in the ground state, and appreciably influences the kinematics of the scattering. The immediate effect is that of allowing higher maximum recoil energy scatterings 
for the same DM mass. This reveals that previously unconsidered light DM masses produce detectable scattering events. In Section~\ref{section:Sensitivity Projections} we establish that this more holistic signal calculation therefore leads to improved sensitivity thresholds compared to models considering only the response of  
individual atoms, and how the immediate effect of lower temperature operation would 
be to make this improvement more pronounced. We also demonstrate that when searching in the sub-MeV realm, one encounters the issue of Pauli-blocking, effectively reducing the fiducial mass of the detector's target, making it increasingly difficult to search for DM with lower and lower masses. In the exploration of this light DM landscape with quantum technologies, one can see explicitly how the quantum statistics within the detector play a critical role in this search.

\section{Dark Matter Scattering in a Many-body Target}\label{section:ManybodyScattering}

Direct detection of DM requires the measurement of energy depositions from scattering events in a detector's target. Corpuscular DM with mass $m_\chi$  scattering in a detector with a many-body target medium of volume $V$ and density $\rho_T$, depositing energy $\omega$, has a differential event rate that follows from Fermi's Golden Rule. The DM is taken to have non-relativistic plane-wave eigenstates with momentum $\bs{k}$ and spin $\tau$, $\ket{\bs{k},\tau}$. The target is described by generic many-body energy eigenstates $\ket{i}$ and $\ket{f}$. DM scatters off the Standard Model (SM) matter of the target through an interaction described by an effective Hamiltonian $H_{\text{int.}}$. The general scattering event rate for a many-body quantum target can then be written as
\begin{equation}
\begin{aligned}
\frac{dR^{\text{int.}}}{d\omega}
&=
\frac{\rho_\chi V}{\rho_T m_\chi}
\int d^3v\,f(\mathbf{v})
\sum_{f,\tau,\tau'}
\int\frac{d^3q}{(2\pi)^3}
\\
&\quad\times
\left|
\bra{\bs{k}-\bs{q},\tau';f}
H_{\text{int.}}
\ket{\bs{k},\tau;i}
\right|^2
2\pi\delta(E_f-E_i-\omega_{\bs{q}})
\delta(\omega-\omega_{\bs{q}}).
\end{aligned}
\label{eq:rate1}
\end{equation}
Here $\rho_\chi$ is the local DM density, $\rho_T$ and $V$ are the mass density and volume of the target, and $f(\mathbf{v})$ is the DM velocity distribution.
The Standard Halo Model \cite{baxter2021recommended, lewin1996review,savage2006annual}
is assumed, in which the DM halo is isothermal and spherically symmetric, with an isotropic Maxwell-Boltzmann velocity distribution truncated at the Galactic escape speed. In this work we use recommended astrophysical parameters \cite{baxter2021recommended}. The scattering of DM with velocity $\mathbf{v}$, involving a momentum transfer $\bs{q}$, deposits an energy $\omega_{\bs{q}}=\mathbf{v}\cdot\bs{q}-\bs{q}^2/(2m_\chi)$. This imposes that, in an isotropic medium such as the superfluid $^3$He B-phase, a DM scattering event that deposits energy $\omega$ and momentum $\bs{q}$ into a target medium requires a minimum incoming DM speed 
\begin{equation}
v_{\text{min}} = \frac{|\bs{q}|}{2m_\chi}+\frac{\omega}{|\bs{q}|}.
\end{equation}

In the following, we consider benchmark Hamiltonians corresponding to the momentum- and velocity-independent non-relativistic effective field theory (NREFT) operators $\mathcal{O}_1$ and $\mathcal{O}_4$.

These operators describe momentum-independent interactions between DM and nucleon fields, $\psi_\chi$ and $\psi_N$. As these operators correspond to SI and SD DM-nucleon cross sections, respectively, we label these interactions SI and SD ($\text{int.}\equiv \text{SI}$ or $\text{int.}\equiv \text{SD}$), with the corresponding interaction Hamiltonians
\begin{equation}
H^{\text{SI}} = c_1\left(\overline{\psi}_\chi\mathbbm{1}_\chi\psi_\chi\right)\left(\overline{\psi}_N\mathbbm{1}_N\psi_N\right),
\qquad
H^{\text{SD}} = c_4\left(\overline{\psi}_\chi\bs{S}_\chi\psi_\chi\right)\cdot\left(\overline{\psi}_N\bs{S}_N\psi_N\right).
\end{equation}
where $c_1$ and $c_4$ are the corresponding dimensionful Wilson coefficients, determined by the parameters of the underlying UV theory of DM.

Assuming no entanglement between the DM and the detector, $\ket{\bs{k},\tau;i}=\ket{\bs{k},\tau}\otimes\ket{i}$. The squared matrix element for the scattering off the many-body target factorises into a DM response and target response. This holds for both SI interactions \cite{trickle2020multi,kahn2022searches}, and many SD interactions\footnote{While UV-complete DM models that lead to a mix of SI and SD EFT operators do not necessarily lead to differential cross sections that factorise in this way, this factorisation holds for the SD interactions considered here; see Appendix~\ref{App:A}. Examples include Axialvector-Axialvector coupling, which produces $\mathcal{O}_4$, the standard SD reference interaction, and Pseudoscalar-Pseudoscalar coupling, which produces $\mathcal{O}_6$.}.
While SD operators are not themselves necessarily separable into operators acting on the DM and nuclear state-spaces, the squared matrix element, summed over DM spins, returns the usual DM spin-sum factor and the target response is hence contained in a separate factor. These matrix elements can be evaluated by writing the NREFT operators in terms of $\bs{S}_\chi$ and $\bs{S}_N$, spin operators acting on the DM and target's constituent nuclei respectively, and $\mathbbm{1}_\chi$ and $\mathbbm{1}_N$ are the identity operators in these spin-spaces. We then see that $\mathcal{O}_1 = \mathbbm{1}_\chi \mathbbm{1}_N$ produces a squared matrix element dependent on the density of target,

\begin{equation}
    \sum_{\substack{\text{DM}\\ \text{ spins}}}\left|{\mathcal{M}(\bs{q})}\right|^2\, \propto\, c_1^2\, \big|\bra{f} \tilde{n}(-\bs{q})\ket{i}\big|^2\label{eq:O1fact}.
\end{equation}
Here $\tilde n(-\bs{q})\equiv\sum_{s}\,\tilde n^{ss}(-\bs{q})$ is the number-density operator constructed from the Fourier transform of a 
number-like operator with two spin indices:
\begin{equation}
    \tilde{n}(-\bs{q})^{s's}=\int\frac{d^3p'}{(2\pi)^3}\frac{d^3p}{(2\pi)^3}(2\pi)^3\delta^{(3)}(\bs{p}'-\bs{p}+\bs{q}){a^{s'}_{\bs{p}'}}^{\dagger}a^s_{\bs{p}},\label{eq:number_FT}
\end{equation}
where ${a^{s}_{\bs{p}}}^{\dagger}$ and $a^s_{\bs{p}}$ are the creation and annihilation operators for states of recoiling nuclei with spin $s$ and momentum $\bs{p}$.  
The SD interaction induced by $\mathcal{O}_4 = \bs{S}_\chi \cdot\bs{S}_N$, factorises only once summed over incoming DM spin-states:
\begin{equation}
    \sum_{\substack{\text{DM}\\ \text{ spins}}}\left|{\mathcal{M}(\bs{q})}\right|^2\, \propto \,c_4^2\,\langle\bs{S}_\chi^2\rangle \big|\bra{f} \tilde{\bs{S}}(-\bs{q})\ket{i}\big|^2,\label{eq:O4fact}
\end{equation}
where we have defined the medium spin-density operator
\begin{equation}\tilde{\mathbf S}(-\bs{q})\equiv\sum_{s's}\mathbf S_N^{s's}\,\tilde n^{s's}(-\bs{q}).\end{equation}

The differential event rate for scattering off the many-body target \eqref{eq:rate1} can then be written in terms of the combined DM response and a dynamic structure function $S^{\text{int.}}(\bs{q},\omega)$, describing the target response:
\begin{equation}
\frac{dR^{\text{int.}}}{d\omega} = \frac{\rho_\chi}{\rho_T m_\chi}\frac{\pi \sigma^{\text{int.}}_{\chi N}}{\mu^2_{\chi N}}\int d^3v\,f(\mathbf{v})\int\frac{d^3q}{(2\pi)^3}S^{\text{int.}}(\bs{q},\omega_{\bs{q}})\delta(\omega- \omega_{\bs{q}}),
    \label{Rate-S}
\end{equation}
where parameters specific to the DM model are contained within the DM-nucleus cross section $\sigma^{\text{int.}}_{\chi N}$.

The parameters of a UV-complete theory of DM can be projected onto the Wilson coefficients in an effective Hamiltonian, from these coefficients we construct the standard reference DM-nucleon cross sections~\cite{ bednyakov2006nuclear} as in Ref.~\cite{darvishi2025dark}: \mbox{$ \sigma^{\text{SI}}_{\chi p}= \left(\mu^2_{\chi p}/\pi\right)\left(c^p_1\right)^2$}, \mbox{$\sigma^{\text{SD}}_{\chi n}=\left[j_\chi(j_\chi+1)\mu^2_{\chi n}/4\pi\right]\left(c^n_4\right)^2$}, with DM of angular momentum $j_\chi$. These nucleon level cross sections allow the direct comparison of experimental constraints to a wealth of DM models. We therefore write the DM-nucleus cross section $\sigma^{\text{int.}}_{\chi N}$ in terms of these reference DM-nucleon cross sections.

The SI DM-nucleus cross section including DM interaction details can be written~\cite{kopp2010global} 
\begin{equation}\sigma^\text{SI}_{\chi N}=\frac{(Zf_p+(A-Z)f_n)^2}{f^2_p}\frac{\mu^2_{\chi_N}}{\mu^2_{\chi p}}\sigma^{\text{SI}}_{\chi p},
\end{equation}
where $\mu_{\chi \,p/n}$ and $\mu_{\chi N}$ are the DM-nucleon and DM-nucleus reduced masses. Here $f_{p}$ and $f_n$ are the DM-proton and DM-neutron couplings. We will assume $f_p=f_n$ such that only the atomic mass number, $A=3$, of the target nucleus is relevant.

Considering the SD interaction, $^3$He has no aggregate average proton spin, $\langle \bs{S}_p\rangle=0$. The SD DM-nucleus cross section therefore only relates to the cross section for the SD DM-neutron interaction:
\begin{align} 
\sigma^{\text{SD}}_{\chi N}&=\frac{\mu^2_{\chi_N}}{\mu^2_{\chi n}}\sigma^{\text{SD}}_{\chi n}\left[\frac{4}{3}\frac{J_N+1}{J_N}\langle\bs{S}\rangle^2\right],\label{eq:SDchiNxsec}
\end{align}
where in $^3$He the total nuclear spin $J_N$ is equal to the mean spin of the neutron, $\langle \bs{S}_n\rangle=\tfrac{1}{2}$.

The dynamic structure function $S^{\text{int.}}(\bs{q},\omega)$ captures the response of the target medium to a scattering event that deposits energy $\omega$ and momentum $\bs{q}$ \cite{trickle2020multi, mazzanti1999finite, kahn2022searches,gori2025spin,schutz2016detectability}.

Following Ref.~\cite{kahn2022searches}, we extend the definition of this function 
to include the action of an interaction specific operator that acts on the target medium's degrees of freedom. Below we define two dynamic structure functions that correspond to the response of the target medium to benchmark SI and SD DM interactions. 

SI DM interactions are diagonal in the spin-space of the target, hence produce a coupling to the density operator. 
The standard dynamic structure function in the SI case then reads
\begin{equation}
    S^{\text{SI}}(\bs{q},\omega)
    =\frac{1}{V}\sum_f \left|\bra{f}\sum_s \tilde{n}(-\bs{q})^{ss}\ket{i}\right|^2 2\pi \delta(E_f-E_i -\omega),\label{eq:SISF}
\end{equation}
where the spin indices have been kept explicit to show how the scattering amplitude is constructed when the target has a non-trivial spin structure.

We also define an SD structure function using the spin-density operator, capturing the interplay between the target spin structure and SD DM interactions. The reference cross section \eqref{eq:SDchiNxsec} is defined so as to include a spin-average appropriate for the scattering response of an individual nucleus. The spin structure of the many-body target medium must be resolved in the structure function and the spin-sum may not simply factorise out. To prevent accounting for the spin response twice, we define the structure function such that it is normalised appropriately; for the SD structure function corresponding to the NREFT operator $\mathcal{O}_4$ we divide through by the spin-averaging factor of a lone nucleus present in the standard SD reference cross section, $s_N(s_N+1)$. The SD structure function then reads
\begin{equation}
    S^{\text{SD}}(\bs{q},\omega)=\frac{1}{V}\sum_f \left|\frac{\bra{f}\tilde{\bs{S}}(-\bs{q})\ket{i}}{\sqrt{s_N(s_N+1)}}\right|^2 2\pi \delta(E_f-E_i -\omega),\label{eq:SDSF}
\end{equation} 
where for spin-$\tfrac{1}{2}$ $^3$He, $s_N(s_N+1)=\tfrac{3}{4}$ 
and the spin operator $\bs{S}_N = \tfrac{1}{2}\boldsymbol{\sigma}$, where $\boldsymbol{\sigma}$ is the vector of Pauli matrices. 
The forms of these structure functions are justified in Appendix~\ref{App:A}. 
This structure function is necessary to describe the low-momentum response of the target where it must be treated as a composite quantum system with long-range coherence. It captures a genuine many-body response and is not reducible to sums over incoherent individual nuclear cross sections. The SD response is
governed by spin-density correlators of the condensate rather than the nuclear spin structure functions, such as $S_{pp}$, $S_{nn}$, $S_{pn}$, that arise in SD scattering from composite nuclei (see Ref.~\cite{quest2024quest,darvishi2025dark}). These deviate from their zero-momentum values only for $q$ above the inverse nuclear radius. This is deep into the incoherent regime where the usual single-atom response is recovered and the individual nucleus spin-sums can be performed as usual. Due to the large separation in length scales between the nuclear scale ($\sim\si{fm}$) and that of the target medium's coherence ($\sim\si{nm}$), in the regime where the target medium is coherent these form factors will be unity, and when the form factors are non-trivial the target is incoherent. One is able to simply join these approaches piecewise at an intermediate energy scales without issue.

Using these dynamic structure functions, one can calculate the differential event rates for both SI and SD interactions. 
In an isotropic target medium the angular component of the momentum integral can be performed. This enforces energy conservation and leaves the rate as an integral over the three dimensions of velocity and the magnitude of transferred 3-momentum $q$. The SI and SD differential event rates are found to be 
\begin{align}
    \frac{dR^{\text{SI}}}{d\omega}&=\frac{9 \rho_\chi \sigma^{\text{SI}}_{\chi \text{p}}}{2 m_\chi \mu^2_{\chi \text{p}}}\frac{2\pi}{\rho_T}
    \int_0^\infty \int^\infty_{v_{\text{min}}} dq\,d\mathbf{v} \,\frac{q}{v}\, f(\mathbf{v}) \,S^{\text{SI}}(q,\omega),\label{isotropicrateSI}\\
    \frac{dR^{\text{SD}}}{d\omega}&=\frac{ \rho_\chi \sigma^{\text{SD}}_{\chi \text{n}}}{2 m_\chi \mu^2_{\chi \text{n}}}\frac{2\pi}{\rho_T}\int_0^\infty \int^\infty_{v_{\text{min}}} dq\,d\mathbf{v} \,\frac{q}{v}\, f(\mathbf{v}) \,S^{\text{SD}}(q,\omega) \label{isotropicrateSD}.
\end{align}
Using the Maxwell-Boltzmann distribution of the Standard Halo Model allows the velocity integral to be performed analytically. The rate is hence reduced to an integral over a single dimension. 

In a semi-classical consideration of the target as a classical gas (discussed in detail in Ref.~\cite{trickle2020multi}) both our SI and SD structure functions, $S^{\text{SI}}(q,\omega)$ \eqref{eq:SISF} and $S^{\text{SD}}(q,\omega)$ \eqref{eq:SDSF}, reduce to the same classical-gas structure function:
\begin{equation}
    S_\text{cl.}(q, \omega)=2\pi \frac{\rho_T}{m_N}\delta\left(\frac{q^2}{2m_N}-\omega\right).\label{eq:SFcl}
\end{equation}
Defining $E_{\text{NR}}=q^2/2m_N$ and $\bar{v}_{\text{min}}(q)\equiv v_{\text{min}}(q,E_{\text{NR}})$, the scattering rates produced from this approximation then correspond to the canonical single-atom DM scattering rates,
\begin{align}
\frac{dR^{\text{SI}}_{\text{cl.}}}{d E_{\text{NR}}}&=
 \frac{ 9\rho_\chi \sigma^{\text{SI}}_{\chi \text{p}}}{2 m_\chi \mu^2_{\chi \text{p}}}\int^\infty_{\bar{v}_{\text{min}}} \,d\mathbf{v} \,\frac{1}{v}\, f(\mathbf{v}),\\
\frac{dR^{\text{SD}}_{\text{cl.}}}{d E_{\text{NR}}}&=
 \frac{ \rho_\chi \sigma^{\text{SD}}_{\chi \text{n}}}{2 m_\chi \mu^2_{\chi \text{n}}}\int^\infty_{\bar{v}_{\text{min}}} \,d\mathbf{v} \,\frac{1}{v}\, f(\mathbf{v}),
 \end{align}
as used in previous work \cite{quest2024quest,darvishi2025attenuation,darvishi2025dark}.

\section{Scattering in Superfluid Helium-3}\label{section:ScatteringInSF3He}

In this section we consider DM scattering in the QUEST-DMC experiment's $^3$He target specifically. The QUEST-DMC detector is operated at temperatures far below $1~\si{mK}$. Under these conditions $^3$He forms a Cooper-pair condensate, a superfluid state. In low momentum scatterings the superfluid $^3$He target must be treated as a composite quantum system with long-range coherence, we hence use the framework above to include relevant many-body effects in the calculation of the detector signal.

While searches for light DM are often restricted to consideration of the DM-electron interactions through electron-counting channels, QUEST-DMC's calorimetric measurement principle, low event-resolution threshold, and the low mass of $^3$He, make it efficient to explore nuclear interactions. We assume elastic scattering such that the entire atom recoils with the nucleus. This is reasonable in low-momentum-transfer scatterings where atomic transitions are incredibly unlikely, or even energetically forbidden. 
For the sub-GeV DM we consider in the analysis below, the ground-state-to-ground-state electron transition amplitude is approximately unity. The lowest energy electronic transition from the atomic ground state is approximately $6\,\si{\eV}$; below this transitions are completely forbidden. We quantise at this atomic level, and hence treat the creation and annihilation operators for the spin and momentum states of the entire atoms, ${a_{\bs{k}}^{s}}^\dagger$, $a_{\bs{k}}^{s}$,
as identical to spin and momentum states of the nuclei. 
As the $^3$He atom's electronic structure is a full 1S shell, this spin corresponds to that of the nucleus, and hence the spin of its lone neutron. Given the electronic contribution to the atomic mass is small, we treat nuclear and atomic mass as interchangeable, as the difference is kinematically negligible. 

Below, the complete scattering off the coherent Bardeen-Cooper-Schrieffer (BCS) Superfluid is considered. This state can be described by a mean-field Hamiltonian. In the mean-field approximation, the pairing interaction is replaced by an order parameter, yielding a quadratic Hamiltonian that couples fermions with opposite momenta to the Cooper-pair condensate. The condensate is described by the gap matrix $\Delta_{\bs{k},ab}$, where $a,b\in\{\uparrow,\downarrow\}$ label the spin-up and spin-down states. In the $^3$He B-phase, this parameter has momentum dependence, which is written using the Pauli matrices as $\Delta_{\bs{k}}=\Delta_0 \left(\hat{\bs{k}}\cdot\boldsymbol{\sigma}\right)i\sigma_2$, where $\Delta_0$ is the magnitude of the superfluid energy gap.
This is diagonalised into a theory of uncorrelated quasiparticles with a Bogoliubov transform, using transformation matrices $U_{\bs{k}}$ and $V_{\bs{k}}$, acting in the same space as the nuclear spin operator $\bs{S}_N$. Due to the lack of spin-orbit coupling $U_{\bs{k}}$ is proportional to the identity in spin space, $V_{\bs{k}}$ captures the nature of the Cooper-channel interaction. We use the conventions 
of Ref.~\cite{vollhardt2013superfluid} where the atomic state creation and annihilation operators are written in terms of the quasiparticle creation and annihilation operators ${\al_{\bs{k}}^{a}}^\dagger,{\al_{\bs{k}}^{a}}$ as follows 
\begin{equation}
    a_{\bs{k}}^{a} = \sum_{b}\left( U_{\bs{k},ab}\al_{\bs{k}}^{b}+V_{\bs{k},ab}{\al_{-\bs{k}}^{b}}^\dagger\right),
\end{equation}
with
\begin{equation}U_{\bs{k},ab}= \delta_{ab}\sqrt{\frac{\xi_{\bs{k}}+E_{\bs{k}}}{2E_{\bs{k}}}},\qquad 
V_{\bs{k},ab} = \frac{\Delta_{\bs{k}, ab}}{\sqrt{2 E_{\bs{k}}(\xi_{\bs{k}}+E_{\bs{k}})}},\label{eq:bgbtrans}\end{equation}
where $m$ is the mass of the $^3$He atom, $\xi_{\bs{k}}=(k^2-k_F^2)/(2m)$ is the free-fermion dispersion relative to the Fermi energy, ${k_F=(3\pi^2n)^{1/3}}$ is the Fermi momentum, and $n$ is the number density. The quasiparticle dispersion is $E_{\bs{k}}=\sqrt{\xi_{\bs{k}}^2+\Delta_0^2}$. The many-body energy eigenstates are then described in terms of quasiparticle excitations of the Cooper-pair condensate.  
The initial and final states we consider are energy eigenstates of the system corresponding to quasiparticles excited from the Cooper-pair condensate.
The ground state is the quasiparticle vacuum, having the condition $\al_{\bs{k}}^{\sigma}\ket{i}=0$  for all $\bs{k}$ and $\sigma$. This is the state of the system at zero-temperature and is a good approximation of the system at the low operating temperatures of the QUEST-DMC detector, see Appendix~\ref{appB}.

The only terms that give a non-trivial amplitude when acting on the quasiparticle vacuum are products of creation operators. The relevant terms from the number-density operator, for SI interactions, are therefore of the form
\begin{equation}
\begin{aligned}
{a^{\uparrow}_{\bs{k}'}}^{\dagger} a^{\uparrow}_{\bs{k}}
\, + \,
{a^{\downarrow}_{\bs{k}'}}^{\dagger} a^{\downarrow}_{\bs{k}}
&\supset
\sum_{a,b} \left(
U^\ast_{\bs{k}',\uparrow a} V_{\bs{k},\uparrow b}
\, + \,
U^\ast_{\bs{k}',\downarrow a} V_{\bs{k},\downarrow b}
\right)
{\alpha^{a}_{\bs{k}'}}^{\dagger}
{\alpha^{b}_{-\bs{k}}}^{\dagger}
\\
&=
({\boldsymbol{\alpha}_{\bs{k}'}}^{\dagger})^{T}
U^\dagger_{\bs{k}'} V_{\bs{k}}
{\boldsymbol{\alpha}_{-\bs{k}}}^{\dagger}.
\end{aligned}
\label{eq:nContainsBogMatrix}
\end{equation}
The final states available to initially scatter into are hence states of exactly two excited quasiparticles with total momentum $\bs{q}$. This determines the immediate kinematics of the scattering. While the superfluid $^3$He is described by a weak-coupling Hamiltonian with non-interacting quasiparticles the physical system eventually thermalises. This allows the measurement of this energy deposition through temperature measurements.

Using Eq.~\eqref{eq:SISF}, we calculate the relevant structure functions. For superfluid $^3$He, since the only non-trivial amplitudes involve creating exactly two quasiparticles, the sum over final states becomes a sum over these quasiparticle spins and the integration is over their momenta, as shown explicitly in Appendix~\ref{App:A}.

Structure functions for SI and SD interactions can be calculated for general spin-pairings. This method therefore also applies directly to DM-electron scattering in superconductors. The structure function for the conventional s-wave pairing superconductor, $\Delta =\Delta_0 (i\sigma_2)$, reproduces the result of Ref.~\cite{hochberg2023directional} in the SI case. 
In the QUEST-DMC detector the target is operated in the isotropic B-phase with a uniform energy gap, this is described by the momentum-dependent order parameter $\Delta_{\bs{k}}=\Delta_0 \left(\hat{\bs{k}}\cdot\boldsymbol{\sigma}\,\,i\sigma_2\right)$.
The nontrivial momentum dependence of the order parameter is relevant, appearing in the cross terms that emerge in the squared amplitude. 

The angular separation of the quasiparticle momenta affects each amplitude, and would be relevant to the differential event rate in an analysis with directional resolution. Here we sum all angular contributions and discriminate events based only on energy deposition.
As one would therefore expect, due to the isotropic order parameter and the isotropic dispersion relations that follow, the sum over all quasiparticle momenta results in isotropic structure functions, independent of $\bs{\hat{q}}$:
\begin{align}
    S^{\text{SI}}(q,\omega)=&\int\frac{d^3k}{(2\pi)^3}\ 2\pi \delta(E_{\bs{q}-\bs{k}}+E_{\bs{k}}-\omega)\times
    \frac{1}{2}\left(1-\frac{\xi_{\bs{q}-\bs{k}}\xi_{\bs{k}}+\Delta_0^2{\left(\bs{q}-\bs{k}\right)\cdot\bs{\hat{k}}\over|\bs{q}-\bs{k}|}}{E_{\bs{q}-\bs{k}}E_{\bs{k}}}\right),\label{eq:SISFSF}\\
    S^{\text{SD}}(q,\omega)=&\int\frac{d^3k}{(2\pi)^3}\ 2\pi \delta(E_{\bs{q}-\bs{k}}+E_{\bs{k}}-\omega)\times
    \frac{1}{2}\left(1-\frac{\xi_{\bs{q}-\bs{k}}\xi_{\bs{k}}+\frac{1}{3}\Delta_0^2{\left(\bs{q}-\bs{k}\right)\cdot\bs{\hat{k}}\over|\bs{q}-\bs{k}|}}{E_{\bs{q}-\bs{k}}E_{\bs{k}}}\right).\label{eq:SDSFSF}
\end{align}
These structure functions describe the complete response of the superfluid to DM-nucleon scattering from the minimum kinematically allowed energy deposition of $2\Delta_0$ to the regime up to energies at which the atom loses coherence.

\begin{figure}[t!]
    \centering
\includegraphics[width=0.8\linewidth]{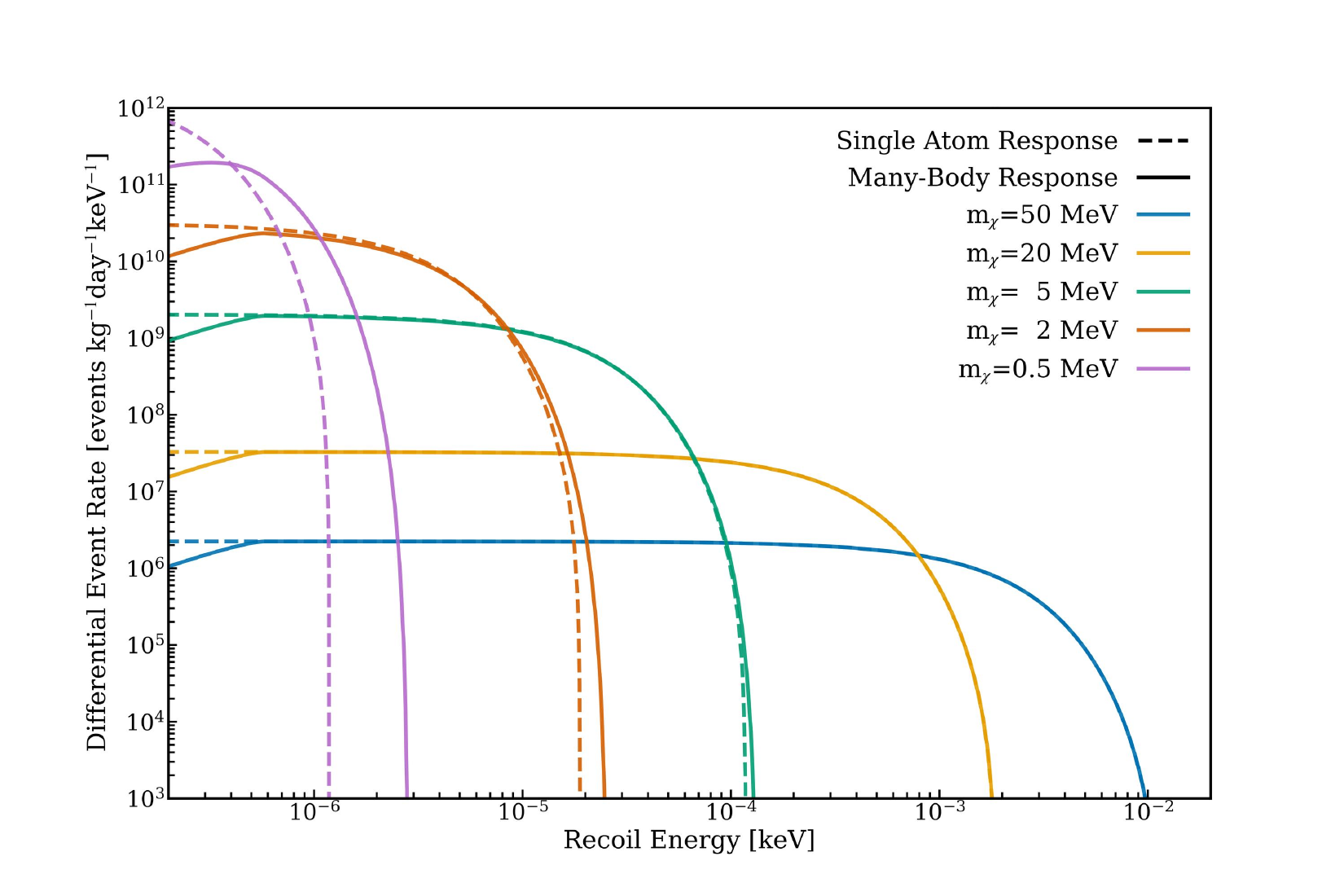}
    \caption{
    SD DM scattering differential event rate (before inclusion of detector effects) for SD DM-neutron cross sections of $10^{-36}\,\si{\square\cm}$, comparing scattering off individual atoms (dashed lines) and the zero-temperature Fermi gas response (solid lines). The shift in the tails is noticeable for DM with masses of a few MeV/c$^2$, where the transferred momentum is close to the Fermi scale.
    In the QUEST-DMC target, Pauli blocking causes the many-body response to drop off at sub-meV recoil energies. For the energy range plotted, the nucleus is coherent and energies are far above the scale of the SD Cooper-channel interaction in the Helium superfluid. The SI rates are therefore identical in form to the SD rates plotted, differing only by a scaling factor $9\mu^2_{\chi \text{n}}/\mu^2_{\chi \text{p}}$ for the cross-section definitions.}
\label{fig:SignalHist}
\end{figure}

The order parameter of the $^3$He superfluid $\Delta_0$ is on the scale $10^{-7}\,\si{\eV}$. This is orders of magnitude below the benchmark $0.51\,\si{\eV}$ threshold for detection of an individual scattering event in the QUEST-DMC detector. For the calculation of the scattering rate used in a standard rare-event search, considering energy depositions above this threshold, we consider the limit where excitations are of energies much higher than the energy gap $\Delta_0$. 
In this limit, $\Delta_0/\xi_{\bs{k}}\rightarrow0$, the effects of the SD pairing interaction vanish. We therefore find that both the SI and SD interactions approach the same limit. The superfluid structure functions (Eqs.~\eqref{eq:SISFSF} and \eqref{eq:SDSFSF}) approach:
\begin{align}
    S^{\text{SI}/\text{SD}}(q,\omega)
    &\rightarrow\int\frac{d^3k}{(2\pi)^3}\ 2\pi \,\delta(\left|\xi_{q-k}\right|+\left|\xi_k\right|-\omega)\times
    \frac{1}{2}\left(1-\frac{\xi_{q-k}\xi_{k}}{\left|\xi_{q-k}\xi_{k}\right|}\right).
\end{align}
As one may expect, at an energy scale away from that of the interatomic interaction, the system is well approximated by a non-interacting Fermi gas description: both the SI and SD expressions are equivalent to the structure function of the zero-temperature Fermi gas (discussed in detail in Ref.~\cite{mazzanti1999finite}) in the limit,
\begin{equation}
   S_{\text{F.G.}}(q,\omega) =\frac{2}{(2\pi)^3}\int^\infty_{\mathbf{k}'=k_F}d^3k' \int^{k_F}_{\mathbf{k}=0}d^3k\, 2\pi\delta(\xi_{k'}-\xi_k-\omega)\delta^{(3)}(\bs{k}'-\bs{k}-\bs{q}).\label{eq:FGSF1}
\end{equation}
In this limit, the coherence factor has become a step function, only allowing scattering of states below a Fermi surface to those above. This factor of two emerges from the two spin species, which in this limit are non-interacting.

Upon cooling, the $^3$He system has undergone a phase transition from a Fermi liquid to a Cooper-pair condensate, corresponding to a ground state distinct from that of the liquid. By taking the limit of the full structure function of the superfluid, the Fermi gas approximation is shown to be an appropriate model for the superfluid scattering for both SI and SD DM interactions in the current region of interest of QUEST-DMC's rare-event search, despite this distinct initial state. Due to the differing ground states and degrees of freedom of the two systems, at finite temperature the results do differ and the correspondence does not apply. At the operating temperatures of the QUEST-DMC experiment, however, the finite-temperature effects on the scattering rate are negligible, see Appendix~\ref{appB}, and the consideration of the system as a Fermi gas provides an analytic form of the scattering for the $^3$He target which provides an accurate description of the signal in the current and near-future regions of interest. The Fermi-gas structure function in Eq.~\eqref{eq:FGSF1} can be evaluated to give the well-known result:
\begin{equation}
S_{\text{F.G.}}(q,\omega) 
    =\left\{\begin{array}{ll}
      \frac{m^2}{\pi q}\omega & q\leq2k_F,\,\frac{2q k_F-q^2}{2m}\geq \omega\geq0, \\
      \frac{m^2}{\pi q}\left(\epsilon_F -\frac{1}{8m}\left(\frac{2m\omega-q^2}{q}\right)^2\right) &q\leq2k_F,\, \frac{2q k_F+q^2}{2m}\geq \omega\geq \frac{2q k_F-q^2}{2m}, \\
      \frac{m^2}{\pi q}\left(\epsilon_F -\frac{1}{8m}\left(\frac{2m\omega-q^2}{q}\right)^2\right)& q\geq2k_F,\,\frac{2q k_F+q^2}{2m}\geq \omega\geq\frac{q^2-2qk_F}{2m},\\
      0&\text{otherwise.}
\end{array} \right.
\end{equation}
This captures the behaviour of fermionic excitations at scales above and below the Fermi momentum. Particles and holes with small momenta are effectively restricted to movement below a Fermi velocity, $\omega\lesssim \tfrac{k_F}{m} q$. For higher momentum responses, excitation of hole states is restricted by the finite depth of the Fermi sphere.  At large momenta the depth of the Fermi sphere is negligible and the response is dominated by the recoil of a particle with approximately free, quadratic dispersion.

\begin{figure}[!t]
    \centering \includegraphics[width=1.01\linewidth]{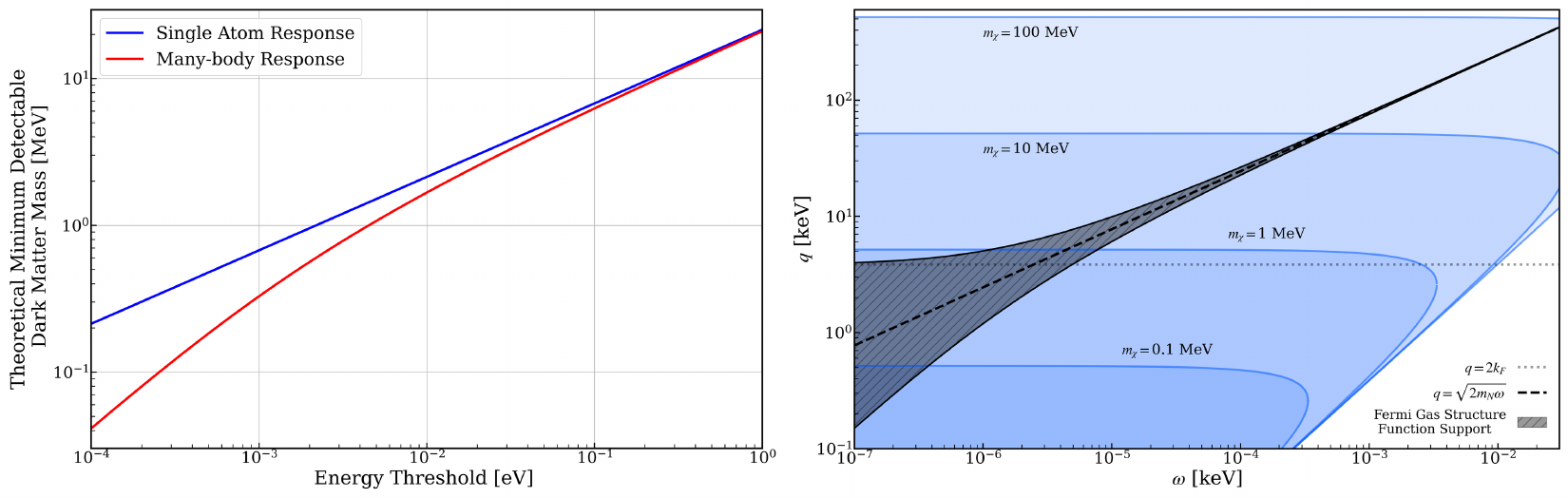}
    \caption{(left) The minimum DM mass that can deposit energy above the energy threshold for $v_{\text{esc}}=544\,\si{\kilo\m\s^{-1}}$ is plotted for the dispersion relation of the classical gas and for the gas of coherent fermions. (right) The accessible region of $(q,\omega)$-space from the DM velocity distribution, kinematically limited by the truncation of the DM velocity distribution at galactic escape velocity, is compared to the single-atom dispersion relation and to the support of the zero-temperature structure function of the Fermi gas at a number density corresponding to $0.15\,\si{g}$ of $^3$He per $\si{\cm ^3}$ for various DM masses. The maximum energy that can be deposited by DM with MeV-scale masses clearly increases by a significant factor when considering the many-body physics of the $^3$He ensemble. 
    }
    \label{fig:accessible}
\end{figure}

The zero-temperature dynamic structure function of the Fermi gas is used to evaluate the differential scattering rate, which is plotted for DM masses 0.5-50 MeV/c$^2$ in Fig.~\ref{fig:SignalHist}. The quantum coherence of the target is seen to affect this rate in two key ways:
\begin{enumerate} 
    \item \emph{Pauli-blocking causes the differential rate to diminish with decreasing energy below the scale of the Fermi momentum.} States within the Fermi sphere are Pauli-blocked from transitions into already occupied states within the Fermi sphere. At a momentum transfer less than the diameter of the Fermi sphere, $q<2k_F$, the effective size of the target is hence decreased, and the magnitude of the signal therefore decreases. This effect only becomes relevant at scales close to and below the Fermi level. The energies of corresponding recoils are currently below the threshold of the QUEST-DMC detector, but would need to be considered if future experiments could be run with a sub-$\si{\milli\eV}$ threshold, relevant to DM with masses a fraction of an MeV/c$^2$.
    \item \emph{The tails of signals for DM with masses of a few MeV/c$^2$ extend to noticeably higher energies compared to those calculated for a non-coherent target.} Without consideration of the target's coherence the dispersion relation of a scattered particle is that of a single free particle, $E_f-E_i = q^2/2m$. Due to Pauli exclusion, the scattering off the Fermi gas involves the scattering of a particle from within a finite Fermi sphere, hence with some finite initial momentum $\bs{k}$, giving many possible energy depositions for a given $\bs{q}$, ${E_f-E_i = (q^2+2\bs{k}\cdot\bs{q})/2m}$; a finite region of $(q,\omega)$-space is accessible for scattering, Fig.~\ref{fig:accessible}. This can also be considered as arising from the extra degree of freedom that comes from creation of two quasiparticles: the excited real particle and the hole in the Fermi sea. 

    The energy cut-off of the signal is given by the truncation of the DM velocity distribution's tail at the galactic escape velocity. The scattering kinematics dictate a minimum DM mass that can deposit energy above a threshold $E_\text{thr}$. 
    For the sub-$\si{\GeV}$ DM considered here $\mu_{N\chi}\simeq m_\chi$, and the approximate threshold DM mass is given by \mbox{$2m_\chi v_\text{esc} = \sqrt{2m E_\text{thr}}$} for the classical target, and \mbox{$2m_\chi v_\text{esc} \simeq \sqrt{2m E_\text{thr}+k_F^2}-k_F$} for the coherent Fermi gas. The consideration of the target's coherence allows sensitivity to lower DM masses for the same energy threshold, Fig.~\ref{fig:accessible}.
\end{enumerate}

This Fermi gas model is appropriate in the regime where DM with few-MeV masses begin to resolve the Fermi scale. For DM of higher masses this model still provides an accurate description of the target response, which closely resembles the single-atom response. We can see this by considering the classical limit $q\gg k_F$. For a given momentum transfer $q>2k_F$, the Fermi-gas structure function restricts the energy to
\begin{equation}
E_{\mathrm{NR}}\left(1-\frac{2k_F}{q}\right)
<\omega<
E_{\mathrm{NR}}\left(1+\frac{2k_F}{q}\right),
\qquad
E_{\mathrm{NR}}\equiv\frac{q^2}{2m}.
\end{equation}
Thus, in the limit $q\gg k_F$, the scattering energy approaches the single-particle dispersion relation,
\begin{equation}
\omega=E_{\mathrm{NR}}
\left[1+\mathcal{O}\left(\frac{k_F}{q}\right)\right].
\end{equation}
When integrated against slowly varying functions, the Fermi-gas structure function therefore approaches the Dirac-delta response of the classical gas given in Eq.~\eqref{eq:SFcl}. 
Considering the event rate differential in $E_{\text{NR}}$, in the limit
$q\gg k_F$ the minimum velocity varies across the allowed energy interval
only by terms of order $\mathcal{O}(k_F/q)$. We may therefore write
$v_{\text{min}}(q,\omega)\simeq
v_{\text{min}}(q,E_{\text{NR}})\equiv\bar{v}_{\text{min}}(q)$.
The SI differential rate then becomes
\begin{align} 
\frac{dR^{\text{SI}}}{d E_{\text{NR}}} 
&= \frac{9\rho_\chi \sigma^{\text{SI}}_{\chi \text{p}}} {2 m_\chi \mu^2_{\chi \text{p}}} \frac{2\pi m}{\rho_T} \int_0^\infty d\omega \int^\infty_{v_{\text{min}}(q,\omega)} d\mathbf{v}\, \frac{1}{v}\,f(\mathbf{v})\, S_{\text{F.G.}}\left(\sqrt{2mE_{\text{NR}}},\omega\right)
\nonumber\\ 
&\simeq \frac{9\rho_\chi \sigma^{\text{SI}}_{\chi \text{p}}} {2 m_\chi \mu^2_{\chi \text{p}}} \frac{2\pi m}{\rho_T} \int^\infty_{\bar{v}_{\text{min}}} d\mathbf{v}\, \frac{1}{v}\,f(\mathbf{v}) \int_0^\infty d\omega\, S_{\text{F.G.}}\left(\sqrt{2mE_{\text{NR}}},\omega\right) 
\nonumber\\ 
&= \frac{9\rho_\chi \sigma^{\text{SI}}_{\chi \text{p}}} {2 m_\chi \mu^2_{\chi \text{p}}} \int^\infty_{\bar{v}_{\text{min}}} d\mathbf{v}\, \frac{1}{v}\,f(\mathbf{v}), 
\end{align}
recovering the single-atom SI differential rate~\cite{quest2024quest}.

As the SD case differs only by a factor $9\mu^2_{\chi \text{n}}/\mu^2_{\chi \text{p}}$, 
the corresponding result is recovered identically. As the single-atom response is recovered for $q\gg k_F$, with $k_F\sim\text{nm}^{-1}$, while the incoherent response of the nucleus is relevant at the $\mathrm{fm}^{-1}$ scale, 
it is therefore consistent to join this approach piecewise with a scattering rate that accounts for the nuclear response at some intermediate momentum scale.
 
\section{Improved Sensitivities for QUEST-DMC}\label{section:Sensitivity Projections}

\begin{figure}[!t]
    \centering    \includegraphics[width=0.8\linewidth]{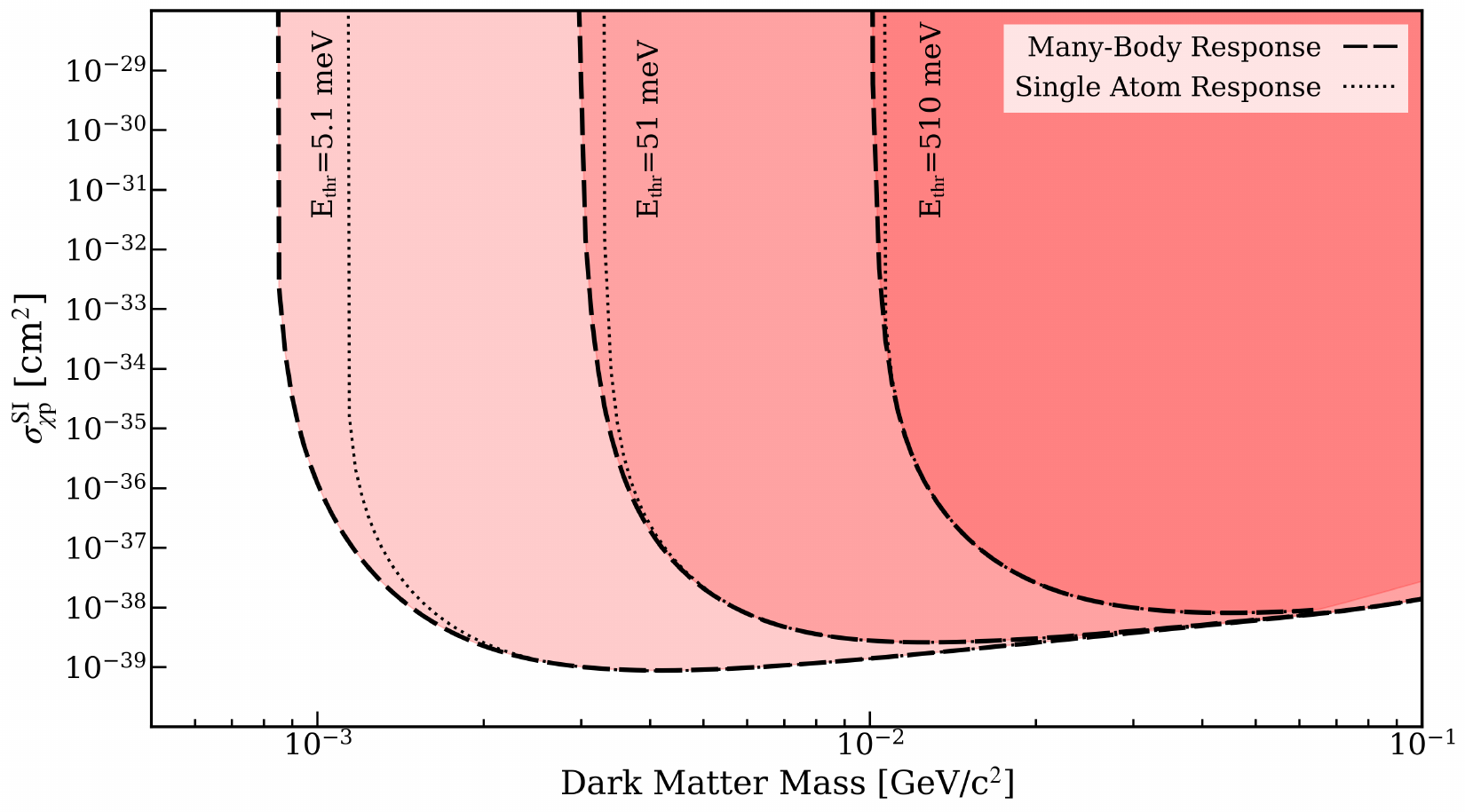}
    \caption{QUEST-DMC sub-GeV DM sensitivities estimated with a Poisson counting analysis, as described in Section \ref{section:Sensitivity Projections}, for three benchmark energy thresholds. It is apparent that the coherence of the detector target affects the mass reach by a discernible fraction for DM with a mass lower than around $10$~MeV/c$^2$. In these projections, the resolution is taken to scale with the detection threshold.
    }
    \label{fig:sens}
\end{figure}

\begin{figure}[!t]
    \centering
\includegraphics[width=1.02\linewidth]{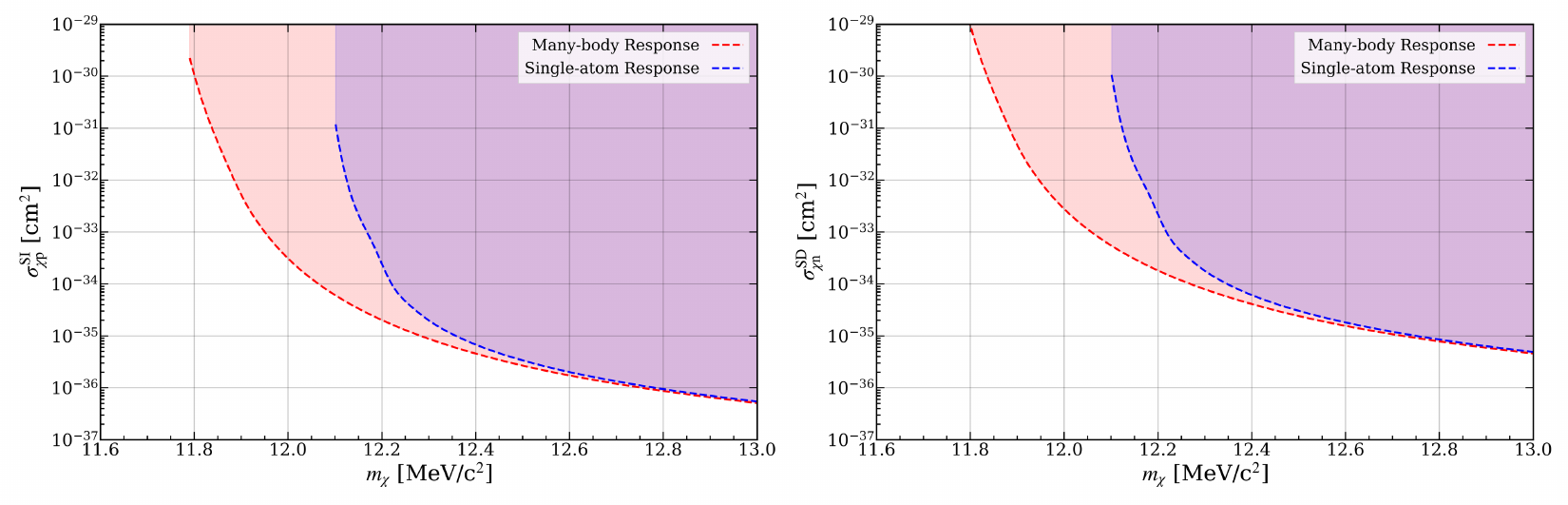}
    \caption{Comparison of QUEST-DMC projected sensitivity for a benchmark threshold of 0.51 eV with SQUID-based readout, over a narrow mass range of a few MeV/c$^2$, produced from a full profile-likelihood-ratio test for both the many-body response and the single-atom approximation.}
    \label{fig:sens_PLR}
\end{figure}

In this section, the sensitivity of the QUEST-DMC experiment to sub-GeV DM is projected by considering a rare-event search under the benchmark detector assumptions adopted here, Fig.~\ref{fig:sens_PLR}, and estimated for potential near-future capabilities, Fig.~\ref{fig:sens}. As the energy threshold of the QUEST-DMC detector is much larger than $\Delta_0$, these sensitivities are projected using DM signals calculated in the 'Fermi gas regime' as discussed above. Sensitivity projections produced in the approximation of an incoherent target, as used in previous work, are made for comparison. The use of this approximation in current projections of QUEST-DMC's reach is validated, while the many-body treatment improves upon it. In future lower-temperature runs, it is apparent that the quantum coherence of the target must be accounted for to accurately describe the mass reach of the experiment.

Detection of DM lighter than 100 MeV/c$^2$ requires the measurement of energy depositions below a few electronvolts. In this region of interest, with a SQUID readout, a dominant error is that of the quasiparticle shot noise, the same effect that determines the detector's energy threshold.

This shot noise is a Poissonian fluctuations from interaction with the thermally excited quasiparticles, whose number scales with temperature as
$$n_{\text{QP}}\propto \sqrt{T}e^{-\Delta/k_{\text{B}}T}.$$
This exponential dependence suggests that modest reductions in operating temperature could substantially lower the energy threshold, provided other noise contributions remain sufficiently small.

A full treatment of the shot-noise involves accounting for the geometry of the nanomechanical resonator, and the dimensions of the bolometer cell, and surface-effects\footnote{We note that the exponential suppression of heat capacity at low temperatures describes the bulk quasiparticle contribution in superfluid $^3$He. Additional contributions may arise from surface-bound quasiparticles and $^3$He adsorbed on the container walls, providing a difficulty in reaching the lower temperatures required for significant changes in threshold. Previous measurements have shown that adsorbed $^3$He can dominate the heat capacity under certain conditions at ultralow temperatures \cite{elbs2007heat}. Preplating with $^4$He can however modify this effect. These effects depend on the surface conditions and are not included in the present bulk description.}.
The energy threshold is set by the ability to discern events above this shot noise, as defined in Ref.~\cite{quest2024quest}. The detector-response effects due to shot noise are applied in two ways. Gaussian smearing due to the shot-noise error is applied to the signal, and a cut is made at the energy threshold.

To appreciate the impact of the improved signal calculation at lower thresholds, a rudimentary Poisson test is carried out in this few-$\si{\eV}$ region of interest relevant to sub-$100$~MeV/c$^2$ DM. The signal is compared to an approximation of the detector's dominant background from radiogenic sources. In this small region of interest, this is taken to be a flat background with a differential rate of $7.76\,\mathrm{events\,kg^{-1}\,day^{-1}\,keV^{-1}}$~\cite{quest2024quest}. In this analysis, the sensitivity is calculated based on a test performed in the optimal region of interest for each mass. In this sensitivity estimate we have assumed the error to be dominated by quasiparticle shot-noise on the resonator, a full analysis would additionally account for electronic noise in the SQUID readout, which could become comparable at these low temperatures.

The projected sensitivity of QUEST-DMC is estimated in this fashion for the benchmark energy thresholds considered here, with 90\% confidence levels shown in Fig.~\ref{fig:sens}.
This analysis demonstrates the impact of the many-body signal model on the exploration of DM parameter space. From Fig.~\ref{fig:sens}, it is clear that at lower energy detection thresholds the single-atom approximation provides a conservative reach, while the more precise many-body treatment of the scattering extends the projected sensitivity to lower DM masses. 
Accounting for the quantum statistics of the target medium, we see that a detector operated with a $5$ meV threshold would be sensitive to sub-MeV DM.

Using the full QUEST-DMC background and error models ~\cite{quest2024quest, autti2024quest} appropriate for an energy threshold of $0.51\,\si{eV}$, a profile-likelihood-ratio test is carried out, as described in Ref.~\cite{quest2024quest}. Here the escape velocity is fixed at a value of $544\,\si{\kilo\m\s^{-1}}$, as recommended in Ref.~\cite{baxter2021recommended}. In Fig.~\ref{fig:sens_PLR}, a comparison is shown between the many-body signal model described above and the single-atom approximation. As was found in the more rudimentary Poisson test, at temperatures of $0.12\,\si{\milli K}$ calculations made in the many-body treatment and in the single-atom approximation agree across almost the entire range of explorable DM masses. Through agreement with the single-atom response model, the use of this approximation in previous sensitivity projections of QUEST-DMC is confirmed quantitatively. The many-body signal model provides an improvement in the mass reach by fractions of an MeV/c$^2$. This demonstrates that the effects of the refined signal model remain discernible even when accounting for parameter uncertainties in a full statistical analysis. In this full treatment of errors and detector effects, the many-body response is necessary to provide an accurate sensitivity to DM with masses of a few MeV/c$^2$ or below. The whole-target scattering model therefore provides a more accurate sensitivity estimate under the benchmark detector assumptions.

\section{Conclusions}

In this paper, we have developed a first-principles description of SI and SD DM scattering in many-body $^{3}$He. We use density and spin-density structure functions to describe the response to the standard momentum- and velocity-independent NREFT operators $\mathcal O_1$ and $\mathcal O_4$, respectively. For the SD interaction, although the scattering amplitude is not generally separable, the differential rate can be factorised into DM and target responses for the class of interactions considered here. We then apply this framework to the B-phase of superfluid $^{3}$He, where scattering from the condensate produces pairs of fermionic quasiparticle excitations.

For energy depositions well above the superfluid gap, the effects of the Cooper-channel interaction become negligible and both the SI and SD structure functions approach the response of a zero-temperature Fermi gas. In the energy range relevant to the rare-event search, this provides an accurate description of the many-body scattering response while retaining the effects of Fermi-Dirac statistics and the finite momentum distribution of the $^{3}$He atoms.

The present calculation represents a step beyond the single-atom approximation by incorporating Fermi statistics and Cooper pairing into the SI and SD responses. We emphasise that, even at energies well above the pairing gap, liquid $^3$He remains a strongly interacting Fermi liquid. A more complete treatment of Fermi-liquid interactions could therefore modify the density and spin-density responses, including through collective excitations, providing a natural direction for future studies of DM scattering in superfluid $^3$He.

The many-body response modifies the scattering kinematics across three characteristic momentum regimes, reflecting the interplay between the quantum statistics of the target and the kinematics of light-DM scattering. At low momentum transfer, Pauli blocking restricts the available final states and suppresses the scattering rate. At momentum transfers comparable to the Fermi momentum, the finite initial momenta of the target particles broaden the range of accessible recoil energies, producing a higher-energy tail that allows lighter DM particles to produce detectable energy depositions. At much larger momentum transfers, $q\gg k_F$, the Fermi-gas structure function approaches the single-atom response. We demonstrate this limit explicitly, providing a first-principles justification of the single-atom approximation in the regime where many-body effects become negligible.

The experimental consequence is most directly reflected in the low-mass reach of QUEST-DMC. At the conventional readout threshold of approximately $31\,\mathrm{eV}$, the effect on the projected mass reach is negligible. At the benchmark SQUID-based readout threshold of $0.51\,\mathrm{eV}$, the full profile-likelihood-ratio analysis shows that the many-body and single-atom calculations agree across almost the entire accessible mass range, quantitatively validating the single-atom approximation underlying previous sensitivity projections. At the lower edge of this range, however, the many-body treatment predicts a higher-energy recoil tail arising from the finite initial momenta of the target atoms. This allows lighter DM particles to deposit sufficient energy to exceed the detector threshold, improving the projected low-mass reach by approximately $2.5\%$ relative to the single-atom approximation. Although the improvement is modest at this benchmark threshold, it becomes increasingly relevant as the detector threshold is lowered, highlighting the importance of accounting for the quantum statistics of the target when approaching the low-mass sensitivity limit.

As lower recoil energies become experimentally accessible, momentum transfers approach the Fermi scale and the difference between the single-atom and many-body descriptions becomes increasingly pronounced. The finite momentum distribution of the target can extend the sensitivity towards progressively lighter DM, with the projected reach ultimately entering the sub-MeV mass range, while at still smaller momentum transfers Pauli blocking restricts the available scattering phase space. Since the Fermi momentum scales with the cube root of the number density, future operation at higher $^{3}$He densities could shift the Fermi scale towards the momentum transfers associated with higher-mass DM, providing an additional route to exploiting the many-body response even without lowering the detector threshold.

The structure-function framework developed here also provides a basis for extending the signal calculation beyond resolved rare-event searches. As a calorimeter, QUEST-DMC could in principle be sensitive to DM energy deposition below the individual-event resolution threshold through its contribution to the detector's baseline heating. Such an analysis would require an accurate description of the low-energy many-body response, for which the structure-function calculation can be extended to finite temperature. A consistent treatment of the many-body response will therefore be important both for determining the ultimate low-mass reach of $^{3}$He-based DM searches and for exploiting future detection strategies below the resolved-event threshold.

\bibliographystyle{JHEP}
\bibliography{ref}

\appendix
\section{Spin-Independent and Spin-Dependent Matrix Elements}\label{App:A}
We consider the scattering matrix element for DM-nucleon interactions described within NREFT \cite{Fan_2010,Fitzpatrick_2013}.  
For a given NREFT operator $\mathcal{O}_i$, the corresponding interaction can be written as
\begin{equation}
    \mathcal{L}_i \supset
    c_i\,
    \overline{\psi}_\chi \mathcal{O}_{\chi,i} \psi_\chi\,
    \overline{\psi}_N \mathcal{O}_{N,i} \psi_N ,
\end{equation}
where $c_i$ denotes the corresponding interaction coefficient. For the momentum- and velocity-independent SI and SD interactions, the corresponding interaction terms are
\begin{equation}
    \mathcal{L}_1 \supset c_1\,\overline{\psi}_\chi \mathbbm{1}_\chi \psi_\chi\,\overline{\psi}_N \mathbbm{1}_N \psi_N,
    \qquad 
    \mathcal{L}_4 \supset
    c_4\, \overline{\psi}_\chi \bs{S}_\chi \psi_\chi
    \cdot
    \overline{\psi}_N \bs{S}_N \psi_N.
\end{equation}
These operators have trivial DM form factors and therefore provide the appropriate interactions for comparison with the standard SI and SD reference cross sections used in previous work \cite{quest2024quest,darvishi2025dark,darvishi2025attenuation}. We use these interactions below to derive the corresponding density and spin-density response functions.

The operator's action on the spin space is made explicit using the corresponding spinors. For spin-$\tfrac{1}{2}$ particles, in the chiral basis the relativistic spinor may be written as
\begin{equation}
    u^s_p =
    \left(
    \sqrt{p\cdot\sigma}\,\xi^s,\,
    \sqrt{p\cdot\bar{\sigma}}\,\xi^s
    \right),
    \qquad 
    S^i=\frac{1}{2}\mathbbm{1}_2\otimes\sigma^i.
\end{equation}

In the non-relativistic limit, while maintaining the same normalisation, we may take $u^s_p=\sqrt{2m}\,\xi^s$ and $S^i=\frac{1}{2}\sigma^i$.
Accordingly, the field expansion reduces to
\begin{equation}
\psi \supset \int \frac{d^3p}{(2\pi)^3}
    \frac{e^{ipx}}{\sqrt{2E_p}} \sum_{s=1,2}u^s_p a^s_p
    \xrightarrow{\bs{p}/m\rightarrow 0} \int \frac{d^3p}{(2\pi)^3} e^{ipx}
    \sum_{s=1,2}\xi^s a^s_p.
\end{equation}
In this limit, $\bar{u}^s_p={u^s_p}^\dagger$.

We assume elastic scattering such that the entire atom recoils with the nucleus. Assuming no entanglement between the DM and the detector, $\ket{\bs{k},\tau;i}=\ket{\bs{k},\tau}\otimes\ket{i}$.
The squared matrix element for this scattering, summed over DM spins, is calculated using unit normalised states.
 Taking the non-relativistic limit and treating the squared energy-conserving delta function in the $T\rightarrow\infty$ limit gives Fermi's Golden Rule,

\begin{align}
\sum_{\tau',\tau}
\left|
\bra{\bs{k}',\tau';f}
\int d^4x\,\mathcal{L}_i
\ket{\bs{k},\tau;i}
\right|^2
&\xrightarrow{p_i/m_N\rightarrow 0}
\sum_{\tau',\tau}
\left|
c_i\,{\mathcal O}^{\tau'\tau}_{\chi,i}
\sum_{s,s'}
{\mathcal O}^{s's}_{N,i}
\bra{f}n(-q)^{s's}\ket{i}
\right|^2
\frac{2\pi\delta(E_f-E_i-\omega_q)T}{V^2}.
\end{align}
For $\mathcal{O}_1$, the SI of the operator results in a response purely dependent on the number density of constituent particles, diagonal in spin space,
\begin{align}
\sum_{\tau',\tau}
\left|\bra{\bs{k}',\tau';f}
\int d^4x\,\mathcal{L}_1
\ket{\bs{k},\tau;i}\right|^2
&= \, 2c_1^2 \, \bigg|\bra{f}n(-q)^{\uparrow\uparrow}+n(-q)^{\downarrow\downarrow}\ket{i} \bigg|^2 \frac{2\pi\delta(E_f-E_i-\omega_q)T}{V^2}.
\end{align}
This leads to Eq.~\eqref{eq:SISF}. 
For $\mathcal{O}_4$, the trace orthogonality of the Pauli matrices allows the spin-summed squared matrix element to factorise. For spin-$\tfrac{1}{2}$ DM,

\begin{align}
\sum_{\tau',\tau}
\left|
\bra{\bs{k}',\tau';f}
\int d^4x\,\mathcal{L}_4
\ket{\bs{k},\tau;i}
\right|^2
&=
\frac{c_4^2}{4}
\sum_{s,s',r,r'}
\sum_{i,j}
\mathrm{Tr}\!\left[\sigma_i\sigma_j\right]
{S_N^{i}}_{s's}
{S_N^{j}}_{rr'}
\nonumber\\
&\quad\times
\bra{f}n(-q)^{s's}\ket{i}
\bra{i}n(q)^{rr'}\ket{f}
\frac{2\pi\delta(E_f-E_i-\omega_q)T}{V^2}
\nonumber\\
&=
\frac{c_4^2}{2}
\left|
\sum_{s,s'\in\{\uparrow,\downarrow\}}
{\bs S_N}_{s's}
\bra{f}n(-q)^{s's}\ket{i}
\right|^2
\frac{2\pi\delta(E_f-E_i-\omega_q)T}{V^2}.
\label{eq:using the trace orthogonality of the Pauli matrices}
\end{align}
This holds for DM with arbitrary spin due to the general trace orthogonality property of spin matrices with orthogonal directions. If the initial and final states are taken to be those of a single nucleus it is clear that a factor $s_N(s_N+1)$ is recovered. This is the single-atom response accounted for in the reference cross section. Normalising by this factor leads to Eq.~\eqref{eq:SDSF}.

It can be seen how working in this framework one could construct the response functions for operators that follow from other theories of DM, such as those resulting in a pseudoscalar nuclear current coupled to a pseudoscalar DM current (upon integrating out a heavy mediator):
\begin{equation}
\mathcal{O}_6 = \left(\frac{\mathbf{q}}{m_N}\cdot \bs{S}_\chi\right)\left(\frac{\mathbf{q}}{m_N}\cdot \bs{S}_N\right),
\end{equation}
with corresponding interaction coefficient $c_6$. For this operator, the response can be written as
\begin{gather}
|F_{\rm DM}(\bs q)|^2 S(\bs{q}, \omega) = \left(\frac{\bs{q}^2}{\bs{q}_0^2}\right)\times \,\frac{1}{V}\sum_f \left|\frac{\bra{f}\bs{q}\cdot\tilde{\bs{S}}(-\bs{q})\ket{i}}{|\bs{q}_0|\sqrt{s_N(s_N+1)}}\right|^2 2\pi \delta(E_f-E_i -\omega),
\end{gather}
where $\bs{q}_0$ is the momentum at which the reference cross section is evaluated.

Using Eqs.~\eqref{eq:SISF} and \eqref{eq:nContainsBogMatrix}, and summing over the spins and momenta of the two-quasiparticle final states, the SI structure function becomes

\begin{align}
S^{\text{SI}}(\bs{q},\omega)
&=
\int\frac{d^3k'}{(2\pi)^3}
\int\frac{d^3k}{(2\pi)^3}\,
2\pi\delta(E_{\bs{k'}}+E_{\bs{k}}-\omega)
(2\pi)^3\delta^{(3)}(\bs{k}'+\bs{k}-\bs{q})
\nonumber\\
&\quad\times
\mathrm{Tr}\Big[
U_{\bs{k'}}U^\dagger_{\bs{k'}}
V_{-\bs{k}}V^\dagger_{-\bs{k}}
\,+\,
U_{\bs{k}}U^\dagger_{\bs{k}}
V_{-\bs{k'}}V^\dagger_{-\bs{k'}}
\nonumber\\
&\qquad
\,-\,
V^\ast_{-\bs{k'}}U^\dagger_{\bs{k'}}
V_{-\bs{k}}U^T_{\bs{k}}
\,-\,
V^\ast_{-\bs{k}}U^\dagger_{\bs{k}}
V_{-\bs{k'}}U^T_{\bs{k'}}
\Big].
\label{SI-s}
\end{align}
For the B-phase, this evaluates to Eq.~\eqref{eq:SISFSF}.

The SD structure function is obtained similarly,  
\begin{align}
S^{\text{SD}}(\bs{q},\omega)
&=
\int\frac{d^3k'}{(2\pi)^3}
\int\frac{d^3k}{(2\pi)^3}\,
2\pi\delta(E_{\bs{k'}}+E_{\bs{k}}-\omega)
(2\pi)^3\delta^{(3)}(\bs{k}'+\bs{k}-\bs{q})
\nonumber\\
&\quad\times\frac{4}{3}
\mathrm{Tr}\Big[
U_{\bs{k'}}U^\dagger_{\bs{k'}}
\bs{S}_N^i
V_{-\bs{k}}V^\dagger_{-\bs{k}}
{\bs{S}_N}_i
\,+\, U_{\bs{k}}U^\dagger_{\bs{k}}
\bs{S}_N^i
V_{-\bs{k'}}V^\dagger_{-\bs{k'}}
{\bs{S}_N}_i
\nonumber\\
&\qquad
\,-\, V^\ast_{-\bs{k'}}U^\dagger_{\bs{k'}}
\bs{S}_N^i
V_{-\bs{k}}U^T_{\bs{k}}
{\bs{S}^\ast_N}_i
\,-\, V^\ast_{-\bs{k}}U^\dagger_{\bs{k}}
{\bs{S}_N}^i
V_{-\bs{k'}}U^T_{\bs{k'}}
{\bs{S}_N^\ast}_i
\Big].
\label{SD-s}
\end{align}
For the B-phase, this reduces to Eq.~\eqref{eq:SDSFSF}.

\section{Structure Functions at Finite Temperature}\label{appB}
In Section~\ref{section:ScatteringInSF3He}, the QUEST-DMC scattering rate is calculated using the zero-temperature superfluid and Fermi-gas structure functions in Eqs.~\eqref{eq:SISFSF}, \eqref{eq:SDSFSF}, and \eqref{eq:FGSF1}. Here we consider the corresponding finite-temperature corrections and show that they are negligible at the operating temperatures relevant to QUEST-DMC. At finite temperature, the dynamic structure function is obtained by thermally averaging over the initial states~\cite{mazzanti1999finite},
\begin{equation}
S(q,\omega;T)
=
\frac{1}{V}
\sum_{f,i}
\frac{e^{-\beta(E_i-\mu N)}}{\mathcal Z}
\left|
\bra f \tilde n(-q)\ket i
\right|^2
2\pi\delta(E_f-E_i-\omega).
\label{eq:finiteTSF}
\end{equation}
For the Fermi gas, the statistics are governed by the Fermi–Dirac distribution $n(k)$. The structure function can be found by considering Eq.~\eqref{eq:finiteTSF} and working in the grand-canonical ensemble \cite{mazzanti1999finite},
\begin{align}
S_{\rm F.G.}(q,\omega;T)
&=
\frac{2}{(2\pi)^3}
\int d^3k'\,d^3k\,
n(k)\bigl[1-n(k')\bigr]
\nonumber\\
&\qquad \qquad\times
2\pi\delta(\xi_{k'}-\xi_k-\omega)
\delta^{(3)}(\bs k'-\bs k-\bs q)
\nonumber\\[1mm]
&=
\frac{m^2}{\pi q}
\frac{1}{\beta(1-e^{-\beta\omega})}
\ln\!\left[
\frac{
1+z\exp\!\left(
-\frac{\beta}{2m}
\left(\frac{2m\omega-q^2}{2q}\right)^2
\right)
}{
1+z\exp\!\left(
-\frac{\beta}{2m}
\left(\frac{2m\omega+q^2}{2q}\right)^2
\right)
}
\right],
\end{align}
where $z=e^{\beta\mu}$ is the fugacity.

For the BCS state, starting from Eq.~\eqref{eq:finiteTSF} and thermally averaging over the QP states, the pair-creation contribution to the structure function is
\begin{align}
S^{\text{int.}}(q,\omega;T)
&=
\int \frac{d^3k'}{(2\pi)^3} \int d^3k\, 
(1-n(k))(1-n(k'))\,
2\pi\delta(E_{k'}+E_k-\omega)\,
\mathcal{F}(\bs{k},\bs{k}')\,
\delta^{(3)}(\bs{k}'+\bs{k}-\bs{q}),
\end{align}
where $\mathcal{F}(\bs{k},\bs{k}')$ is the coherence factor determined by the specific form of the traces in the structure function. The zero-temperature pair-creation result is recovered when the probability that the two final QP states are initially unoccupied is taken to be unity. This is appropriate as we consider temperatures where fewer than one in ten thousand QPs are thermally excited,
\begin{equation}
T\simeq 0.1\Delta
\quad\Longrightarrow\quad
n(k)\lesssim e^{-9}.
\end{equation}
These low temperatures therefore also correspond to significantly smaller energies than those of the excitations we consider in our Fermi gas approximation, which by definition applies away from the scale of the gap. The Fermi--Dirac distribution, with a width $T\lesssim\Delta$, is therefore very well approximated by the step function used implicitly in Eq.~\eqref{eq:FGSF1}.
\end{document}